\documentclass[12pt]{article}

\usepackage{url, graphicx, color, varioref, amsfonts, longtable, float}
\usepackage[matrix,frame,arrow]{xypic}
\input{Qcircuit}

\newtheorem{theorem}{Theorem}[section]

\newtheorem{definition}[theorem]{Definition}
\newtheorem{observation}[theorem]{Observation}

\newcommand{\qed}{\nobreak \ifvmode \relax \else
      \ifdim\lastskip<1.5em \hskip-\lastskip
      \hskip1.5em plus0em minus0.5em \fi \nobreak
      \vrule height0.75em width0.5em depth0.25em\fi}

\title{Synthesis and Optimization of Reversible Circuits for Homogeneous Boolean Functions}

\author{Ahmed Younes\footnote {A.Younes@sci.alex.edu.eg}\\
Department of Math. \& Comp. Science\\
Faculty of Science\\
Alexandria University\\
Alexandria, Egypt}

\begin{document}
\maketitle
\begin{abstract}
Homogenous Boolean function is an essential part of any
cryptographic system. The ability to construct an optimized
reversible circuits for homogeneous Boolean functions might arise
the possibility of building cryptographic system on novel
computing paradigms such as quantum computers. This paper shows a
factorization algorithm to synthesize such circuits.
\end{abstract}

\section{Introduction}

Reversible logic \cite{bennett73,fredtoff82} is one of the hot areas of research. It has many
applications in quantum computation \cite{Gruska99,nc00a}, low-power CMOS \cite{cmos2,cmos1} and
many more. Synthesis and optimization of reversible circuits cannot be done using conventional ways \cite{toffoli80}.

The design and analysis of Boolean based devices has many
applications in engineering and science, e.g. cryptography.
Massive computation power is required as the complexity and so the
strength of encryption algorithms increase. Secure encryption
algorithms involves designing two elementary blocks: P-box and
S-box \cite{677984}. The S-Box theory is the design and analysis
of Boolean functions that have certain desirable cryptographic
properties such as balance, symmetry and high nonlinearity
\cite{qu00homogeneous}. Synthesis and optimization of Boolean
systems on non-standard computers that promise to do computation
more powerfully \cite{simon94} than classical computers, such as
quantum computers, is an essential aim in the exploration of the
benefits that may be gain from such systems.

A lot of work has been done trying to find an efficient reversible circuit for an arbitrary reversible
function. Reversible truth table can be seen as a permutation matrix of size $2^n\times2^n$.
In one of the research directions, it was shown that the process of synthesizing linear reversible
circuits can be reduced to a row reduction problem of $n \times n$ non-singular matrix \cite{patel}.
Standard row reduction methods such as Gaussian elimination and LU-decomposition have been
proposed \cite{Beth01}. In another research direction, search algorithms and template matching
tools using reversible gates libraries have been used \cite{Dueck,Maslov1,Maslov2,DMMiller2,DMMiller1}.
These will work efficiently for small circuits. Benchmarks for reversible circuits have been
established \cite{Benurl1,Benurl2}.

%Intro for Homogeneous Function

Implementing Boolean functions with $n$ inputs and a single output
as reversible circuits using some of the above methods is not
possible for certain classes of Boolean functions \cite{revlogic},
where the inputs are required to stay unchanged as required by
many quantum algorithms \cite{grover96,Younes03c}, where further
operations will be applied on the inputs based on the output of
the Boolean function . For example, representing the truth table
of a Boolean function as a non-singular $n \times n$ matrix may
not be possible in some cases. Implementing Boolean function as a
reversible function using search algorithms could be unnecessarily
exhaustive, since we can immediately drop half the reversible
truth table and keep track only for the changes to the single
output, since no inputs will be changed. Recently, there have been
few efforts to find methods to create efficient Boolean reversible
circuits. A method was proposed where it used a ROM-based model
such that the inputs might not be changed even during an
intermediate stage \cite{ROM}. Using this model will require an
exponential number of ROM calls. In \cite{Younes03b}, it was shown
that there is a direct correspondence between reversible Boolean
operations and certain forms of classical logic known as
Reed-Muller expansions. This arises the possibility of handling
the problem of synthesis and optimization of reversible Boolean
logic within the field of Reed-Muller logic. In another research
direction \cite{wav}, it was suggested that fixed polarity
Reed-Muller expansions (FPRM) \cite{bookRM} can be used with
binary decision diagrams (BDD) in an iterative algorithm to
generate reversible circuits for simple incompletely specified
Boolean functions with less than 10 variables. A method proposed
in \cite{practmethod} used a modified version of {\it Karnaugh
maps} and depends on a clever choice of certain minterms to be
used in the minimization process. However, this algorithm may have
poor scalability because of the usage of Karnaugh maps. Another
method is given in \cite{transrules}, where a very useful set of
transformations for Boolean quantum circuits was shown. In this
method, extra auxiliary bits are used in the construction that
will increase the hardware cost.

In this paper, a simple algorithm for synthesis and optimization
by factorization of homogeneous Boolean functions of $n$ variables
with degree $\le3$ will be presented. The algorithm highly
optimizes the quantum cost and the number of gates of the
reversible circuit. The structure of the paper is as follows:
Section \ref{sec2} reviews the necessary background. Section
\ref{sec3} presents two methods for the reversible construction of
homogeneous Boolean functions where the experimental results will
be presented. The paper ends up with a conclusion in Section
\ref{sec4}.

\section{Background}
\label{sec2}

\subsection{Reversible Boolean Function}
A Boolean function is a function that takes $n$ Boolean inputs and
generates a single Boolean output.
\begin{definition} (Boolean Function)

Any Boolean function $f$ with $n$ variables $f:\left\{ {0,1}
\right\}^n \to \left\{ {0,1} \right\}$ can be represented as a
Positive Polarity Reed-Muller (PPRM) expansion as follows
\cite{boolth},

\begin{equation}
f(x_{n - 1} ,...,x_1 ,x_0 ) =  \mathop \bigoplus\limits_{i =
0}^{2^n  - 1} {b_i \varphi _i }, \label{eqn2}
\end{equation}

\noindent where,

\begin{equation}
\label{eqn3} \varphi _i = \prod\limits_{k = 0}^{n - 1}
\left({\mathop x\limits}  _{k}\right)^{i_k},
\end{equation}

\noindent where $x_k ,b_i \in \left\{ {0,1} \right\}$ and $i_k$
represent the binary digits of $i$. $\varphi _i$ are known as
product terms (minterms) and $b_i $ determine whether a minterm is
present or not. $\bigoplus$ means that the arguments are subject
to Boolean operation exclusive-$OR$ $(XOR)$ and multiplication is
assumed to be the $AND$ operation.
\end{definition}

For example, the set of all 3-inputs Boolean functions can be
represented as:
\begin{equation}
f(x_2,x_1,x_0) = b_0 \oplus b_1x_0 \oplus b_2x_1 \oplus b_3x_1x_0
\oplus b_4x_2 \oplus b_5x_2x_0 \oplus b_6x_2x_1 \oplus
b_7x_2x_1x_0.
\end{equation}
Setting $b_i$'s to 0 or 1 gives different 3-inputs Boolean
functions.

\subsection{Reversible Boolean Circuits}

In building a reversible Boolean circuit for a given Boolean
function with $n$ variables, an $n+1 \times n+1$ reversible
circuit will be used, where the extra auxiliary bit will be
initialize to zero, to hold the result of the Boolean function at
the end of the computation. $CNOT$ gate is the main primitive gate
that will be used in building the circuit.  $CNOT$ gate is defined
as follows:

\begin{definition}($CNOT$ gate)

$CNOT(x_{n-1}, x_{n-2},\ldots,x_{0};f)$ is a reversible gate with
$n+1$ inputs $x_{n-1}$, $x_{n-2}$,$\ldots,x_{0}$ (known as control
bits) and $f_{in}$ (known as target bit) and $n+1$ outputs
$y_{n-1}$, $y_{n-2}$,$\ldots,y_{0}$ and ${f_{out}}$. The operation
of the $CNOT$ gate is defined as follows,

\begin{equation}
\begin{array}{l}
 y_i  = x_i ,for\,0 \le i \le n - 1, \\
 f_{out}  = f_{in}  \oplus x_{n - 1} x_{n - 2}  \ldots x_0, \\
 \end{array}
\end{equation}
\noindent
\end{definition}
i.e. the target bit will be flipped if and only if all the control
bits are set to 1. Some special cases of the general $CNOT$ gate
have their own names, $CNOT$ gate with no control bits is called
$NOT$ gate (Figure~\ref{figcnot}.b), where the bit which will be
flipped unconditionally. $CNOT$ gate with one control bit is
called {\it Feynman} gate (Figure~\ref{figcnot}.c). $CNOT$ gate
with two control bits is called {\it Toffoli} gate
(Figure~\ref{figcnot}.d).

\begin{center}
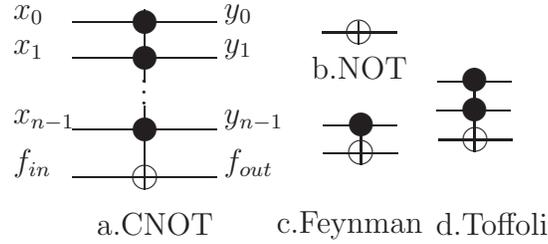
\begin{figure}[t]
\begin{center}
\setlength{\unitlength}{3947sp}%
\begingroup\makeatletter\ifx\SetFigFont\undefined%
\gdef\SetFigFont#1#2#3#4#5{%
  \reset@font\fontsize{#1}{#2pt}%
  \fontfamily{#3}\fontseries{#4}\fontshape{#5}%
  \selectfont}%
\fi\endgroup%
\begin{picture}(3169,1438)(5451,-1937)
{\thinlines
\put(6301,-586){\circle*{150}}
}%
{\put(6301,-811){\circle*{150}}}%
{\put(6301,-1261){\circle*{150}}}%
{\put(6301,-1561){\circle{150}}}%
{\put(7658,-1404){\circle{150}}}%
{\put(7661,-1231){\circle*{150}}}%
{\put(8376,-1322){\circle{150}}}%
{\put(8373,-1136){\circle*{150}}}%
{\put(8372,-949){\circle*{150}}}%
{\put(7645,-649){\circle{150}}}%
{\put(6751,-811){\line(-1, 0){900}}}%
{\put(6751,-1261){\line(-1, 0){900}}}%
{\put(6751,-1561){\line(-1, 0){900}}}%
{\put(6301,-511){\line( 0,-1){450}}}%
\put(6270,-1100){\makebox(0,0)[lb]{{$\vdots$}}}

{\put(6301,-1111){\line( 0,-1){525}}}%
{\put(6751,-586){\line(-1, 0){900}}}%
{\put(7424,-1407){\line( 1, 0){463}}}%
{\put(7659,-1162){\line( 0,-1){318}}}%
{\put(7422,-1233){\line( 1, 0){463}}}%
{\put(8145,-1325){\line( 1, 0){463}}}%
{\put(8372,-877){\line( 0,-1){520}}}%
{\put(8137,-1142){\line( 1, 0){463}}}%
{\put(8140,-959){\line( 1, 0){463}}}%
{\put(7418,-650){\line( 1, 0){463}}}%
{\put(7644,-582){\line( 0,-1){140}}}%

\put(5476,-597){\makebox(0,0)[lb]{{$x_{0}$}}}
\put(5476,-814){\makebox(0,0)[lb]{{$x_{1}$}}}
\put(5476,-1267){\makebox(0,0)[lb]{{$x_{n-1}$}}}
\put(5476,-1561){\makebox(0,0)[lb]{{$f_{in}$}}}

\put(6800,-1561){\makebox(0,0)[lb]{{$f_{out}$}}}
\put(6800,-1267){\makebox(0,0)[lb]{{$y_{n-1}$}}}
\put(6800,-814){\makebox(0,0)[lb]{{$y_{1}$}}}
\put(6800,-597){\makebox(0,0)[lb]{{$y_{0}$}}}

\put(7130,-1950){\makebox(0,0)[lb]{{c.Feynman}}}
\put(8131,-1922){\makebox(0,0)[lb]{{d.Toffoli}}}
\put(7350,-939){\makebox(0,0)[lb]{{b.NOT}}}
\put(6000,-1922){\makebox(0,0)[lb]{{a.CNOT}}}
\end{picture}%

\end{center}
\caption{\label{figcnot}$CNOT$ gates. The back circle $\bullet $
indicates the control bits, and the symbol $ \oplus $ indicates
the target bit. (a)$CNOT$ gate with $n$ control bits. (b) $CNOT$
gate with no control bits. (c)$CNOT$ gate with one control bit.
(d)$CNOT$ gate with two control bits.}
\end{figure}
\end{center}

\subsection{Quantum Cost}

{\it Quantum cost} is a term appears in the literature to refer to
the technological cost of building $CNOT$ gates. The total quantum
cost of a $CNOT$-based reversible circuit is subject to
optimization as well as the number of $CNOT$ gates used in the
circuit. Quantum cost based primarily on the number of control
bits per $CNOT$ gates. Quantum cost refers to the number of
elementary operations required to build the $CNOT$ gate
\cite{PhysRevA.52.3457,1049300}. The calculation of quantum cost
for the circuits in this paper will be based on the cost table
available in \cite{Benurl2}. The state-of-art shows that both
$CNOT(x_i)$ and $CNOT(x_i;f)$ have quantum cost = 1,
$CNOT(x_i,x_j;f)$ has a quantum cost = 5, and
$CNOT(x_i,x_j,x_k;f)$ has a quantum cost = 13. Interaction of such
gates in different ways may give a total quantum cost of the
circuit less than the sum of the quantum cost of each $CNOT$ gate
in the circuit.

\subsection{Homogeneous Boolean Functions}

Homogeneity and nonlinearity are important properties of Boolean
functions when used in cryptographic algorithms. Boolean functions
with the highest possible nonlinearity are called homogeneous bent
functions \cite{677984,qu00homogeneous}.

\begin{definition} (Homogeneous Boolean Function)

A Boolean function $f:\left\{ {0,1} \right\}^n  \to \left\{ {0,1}
\right\}$ is homogeneous of degree $k$ if it can be represented as
follows \cite{GF2n},

\begin{equation}
f(x_{n - 1} , \ldots ,x_{1},x_0 ) = \mathop  \bigoplus \limits_{0
\le i_1  \le  \ldots  \le i_k  \le n - 1} b_{i_1 \ldots i_k }
x_{i_1 }  \ldots x_{i_k},
\end{equation}
\noindent
where each term $x_{i_1 }  \ldots x_{ik}$, $b_{i_1
\ldots i_k }$ is a product of precisely $k$ variables.
\end{definition}

For example, it was found that there are 20 distinct minterms of
degree 3 in a Boolean function of six variables and hence there
are $2^{20}$ possible homogeneous Boolean function of degree 3 on
six variables. Within this space, there are 30 homogeneous bent
functions with 16 minterms. For example, a representative of a
homogeneous bent function of degree 3 on six variables is as
follows \cite{qu00homogeneous},
\begin{equation}
\label{HBFR}
\begin{array}{l}
 f={x}_{ 0} {x}_{ 1} {x}_{2} {  } \oplus {x}_{ 0} {x}_{ 1} {x}_{3} { } \oplus { x}_{ 0} { x}_{ 1} { x}_{ 4} {  } \oplus {  x}_{ 0} { x}_{ 1} { x}_{ 5} {  } \\
  \,\,\,\,\,\,\,\,\oplus {  x}_{ 0} { x}_{ 2} { x}_{ 3} {  } \oplus {  x}_{ 0} { x}_{ 2} { x}_{ 5} {  } \oplus {  x}_{ 0} { x}_{ 3} { x}_{ 4} {  } \oplus {  x}_{ 0} { x}_{ 4} { x}_{ 5} {  } \\
  \,\,\,\,\,\,\,\,\oplus {  x}_{ 1} { x}_{ 2} { x}_{ 3} {  } \oplus {  x}_{ 1} { x}_{ 2} { x}_{ 4} {  } \oplus {  x}_{ 1} { x}_{ 3} { x}_{ 5} {  } \oplus {  x}_{ 1} { x}_{ 4} { x}_{ 5}  \\
  \,\,\,\,\,\,\,\,\oplus {  x}_{ 2} { x}_{ 3} { x}_{ 4} {  } \oplus {  x}_{ 2} { x}_{ 3} { x}_{ 5} {  } \oplus {  x}_{ 2} { x}_{ 4} { x}_{ 5} {  } \oplus {  x}_{ 3} { x}_{ 4} { x}_{ 5}.  \\
 \end{array}
\end{equation}

\section{Synthesis Algorithm}
\label{sec3}
\subsection{Direct Synthesis}

In this section, the steps to implement any arbitrary Boolean
function $f$ using positive polarity RM expansions as reversible
circuits will be presented. For example, consider the homogeneous
function,
\begin{equation}
\label{DSex} f\left( {x_4,x_3,x_2 ,x_1 ,x_0 } \right) = x_2 x_1
x_0 \oplus x_4 x_3 x_0 \oplus x_4 x_3 x_1 \oplus x_4 x_3 x_2
\oplus x_3 x_1 x_0.
\end{equation}
To find the reversible circuit implementation for this function, we
may follow the following procedure:

\begin{itemize}

\item[1-]For a Boolean function with $n$ variables, prepare $n$
bits and initialize an extra bit $t$ to 0 , which will hold the
result of the Boolean function.

\item[2-]Add a $CNOT$ gate for each product term in this expansion taking the Boolean
variables in this product term as control bits and the extra bit as the
target bit.

\item[3-]For the product term, which contains 1, add a $CNOT(t)$,
so the final circuit for $f$ shown in Eqn. \ref{DSex} will be as
shown in Fig. \ref{expol0}.
\end{itemize}

\begin{center}
\begin{figure} [htbp]
\begin{center}
\[
\Qcircuit @C=1em @R=1em {
\lstick{x_0}    & \ctrl{5}  & \ctrl{5}  &\qw    &\qw    & \ctrl{5}  &\rstick{x_0}\qw \\
\lstick{x_1}    & \ctrl{4}  &\qw    & \ctrl{4}  &\qw    & \ctrl{4}  &\rstick{x_1}\qw \\
\lstick{x_2}    & \ctrl{3}  &\qw    &\qw    & \ctrl{3}  &\qw    &\rstick{x_2}\qw \\
\lstick{x_3}    &\qw    & \ctrl{2}  & \ctrl{2}  & \ctrl{2}  & \ctrl{2}  &\rstick{x_3}\qw \\
\lstick{x_4}    &\qw    & \ctrl{1}  & \ctrl{1}  & \ctrl{1}  &\qw    &\rstick{x_4}\qw \\
\lstick{0}  & \targ & \targ & \targ & \targ & \targ &\rstick{f}\qw
}
\]
\end{center}
\caption{Reversible circuit realization using the direct method of
the homogeneous Boolean function shown in Eqn.\ref{DSex}.}
\label{expol0}
\end{figure}
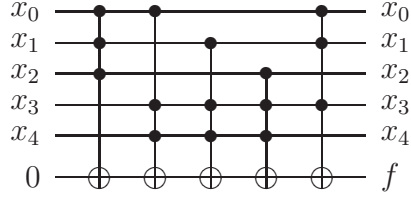
\end{center}

\subsection{Optimization by Factorization}

The main idea of factorization is to find the common terms in a
Boolean function that have high quantum cost and implement them in
a way to decrease multi-calculations of such terms and so decrease
the total quantum cost and/or the total number of $CNOT$ gates.
The following observations are essential in understanding the
final construction:

\begin{observation}
\label{obs1} For a homogeneous Boolean function of degree 3 of
the form,

\begin{equation}
x_i x_j x_{k_1 }  \oplus x_i x_j x_{k_2 }  \oplus  \cdots  \oplus
x_i x_j x_{k_n },
\end{equation}
\noindent
it can be factorized to take the form,
\begin{equation}
x_i x_j \left( {x_{k_1 }  \oplus x_{k_2 }  \oplus  \cdots  \oplus
x_{k_n } } \right).
\end{equation}
\noindent
Realization of that expression as a reversible circuit
is as follows,

\begin{equation}
\begin{array}{l}
 CNOT(x_{k_1 } ;x_{k_2 } )CNOT(x_{k_2 } ;x_{k_3 } ) \ldots (x_{k_{n - 1} } ;x_{k_n } ) \\
 CNOT(x_i ,x_j ,x_{k_n } ;f)CNOT(x_{k_{n - 1} } ;x_{k_n } ) \ldots CNOT(x_{k_2 } ;x_{k_3 } ) \\
 CNOT(x_{k_1 } ;x_{k_2 } ). \\
 \end{array}
\end{equation}

Synthesis of that circuit will decrease the quantum cost from
$13n$ is constructed using the direct method to $13+2(n-1)$ after
factorization. If the reservation of the inputs is not important,
then the quantum cost can be decreased to $12+n$, where the last
$(n-1)$ $CNOT$ gates used to restore the state of the inputs can
be removed. For example, the following function with a circuit of
quantum cost = 39,
\begin{equation}
\label{ob1ex} f=x_0 x_3 x_4  \oplus x_1 x_3 x_4  \oplus  x_2 x_3
x_4,
\end{equation}
\noindent
can be re-written as follows,
\begin{equation}
f=x_3 x_4 \left( {x_0  \oplus x_1  \oplus x_2 } \right),
\end{equation}
\noindent such function has a circuit of quantum cost = 17 as
shown in Fig.\ref{obs1fig}. Note that, the last two $CNOT$ gates
can be removed of the values of the inputs is not important.
\end{observation}

\begin{center}
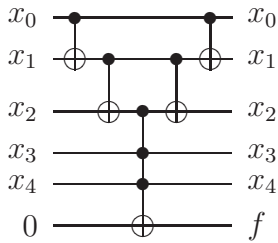
\begin{figure} [htbp]
\begin{center}
\[
\Qcircuit @C=.3em @R=1em {
\lstick{x_0}    & \ctrl{1}  &\qw    &\qw    &\qw    & \ctrl{1}  &\rstick{x_0}\qw \\
\lstick{x_1}    & \targ & \ctrl{1}  &\qw    & \ctrl{1}  & \targ &\rstick{x_1}\qw \\
\lstick{x_2}    &\qw    & \targ & \ctrl{3}  & \targ &\qw    &\rstick{x_2}\qw \\
\lstick{x_3}    &\qw    &\qw    & \ctrl{2}  &\qw    &\qw    &\rstick{x_3}\qw \\
\lstick{x_4}    &\qw    &\qw    & \ctrl{1}  &\qw    &\qw    &\rstick{x_4}\qw \\
\lstick{0}  &\qw    &\qw    & \targ &\qw    &\qw    &\rstick{f}\qw
}
\]
\end{center}
\caption{Reversible circuit realization for the Boolean function
shown in Eqn. \ref{ob1ex} based on Observation \ref{obs1}}
\label{obs1fig}
\end{figure}
\end{center}

\begin{observation}
\label{obs21} If a homogeneous Boolean function can be represented
in the form,
\begin{equation}
\begin{array}{l}
 x_{i_1 } x_{j_1 } \left( {x_{k_0 }  \oplus  \cdots  \oplus x_{k_{n } } } \right) \oplus x_{i_2 } x_{j_2 } \left( {x_{k_0 }  \oplus  \cdots  \oplus x_{k_{n}} \oplus x_{k_{n+1}}} \right) \\
  \oplus  \ldots  \oplus x_{i_m } x_{j_m } \left( {x_{k_0 }  \oplus  \cdots \oplus x_{k_{n}} \oplus \cdots \oplus x_{k_{n+m-1} } } \right), \\
 \end{array}
\end{equation}
\noindent then the upper bound of the quantum cost will be
decreased from $13m\left( {n + \frac{{m + 1}}{2}} \right) =
O\left( {m^2 } \right)$ using the direct method  to $13m+2(n+m-1)=
O\left( {m } \right)$ and the number of $CNOT$ gates will be
$m+2(n+m-1)$ instead of $m\left( {n + \frac{{m + 1}}{2}} \right)$.
For example, the function,
\begin{equation}
\label{obs2ex} f= x_2 x_3 \left( {x_0  \oplus x_1 } \right) \oplus
x_3 x_4 \left( {x_0  \oplus x_1  \oplus x_2 } \right),
\end{equation}
\noindent has a circuit with quantum cost = 30 instead of 65
($m=2,n=1$) if synthesized using the direct method as shown in
Fig.\ref{obs2}. Note that, the last two $CNOT$ gates can be
removed if the values of the inputs is not important. Finding the
maximum number of common factors $m$ leads to a decrease in the
number of $CNOT$ gates as well as the total quantum cost of the
reversible circuit.
\end{observation}

\begin{center}
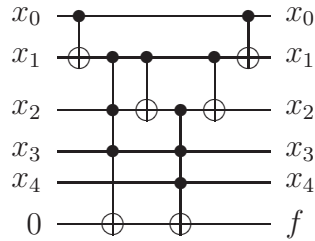
\begin{figure} [htbp]
\begin{center}
\[
\Qcircuit @C=.3em @R=1em {
\lstick{x_0}    & \ctrl{1}  &\qw    &\qw    &\qw    &\qw    & \ctrl{1}  &\rstick{x_0}\qw \\
\lstick{x_1}    & \targ & \ctrl{4}  & \ctrl{1}  &\qw    & \ctrl{1}  & \targ &\rstick{x_1}\qw \\
\lstick{x_2}    &\qw    & \ctrl{3}  & \targ & \ctrl{3}  & \targ &\qw    &\rstick{x_2}\qw \\
\lstick{x_3}    &\qw    & \ctrl{2}  &\qw    & \ctrl{2}  &\qw    &\qw    &\rstick{x_3}\qw \\
\lstick{x_4}    &\qw    &\qw    &\qw    & \ctrl{1}  &\qw    &\qw    &\rstick{x_4}\qw \\
\lstick{0}  &\qw    & \targ &\qw    & \targ &\qw    &\qw    &\rstick{f}\qw
}
\]
\end{center}
\caption{Reversible circuit realization for the Boolean function
shown in Eqn. \ref{obs2ex} based on Observation \ref{obs21}.}
\label{obs2}
\end{figure}
\end{center}
\subsubsection{Factorization Algorithm}

From Observation \ref{obs21}, we can see that finding the maximum
number of common factors in the expression will optimize the
quantum cost and the number of $CNOT$ gates of the reversible
circuit. Similar observations can be constructed for homogeneous
Boolean functions of degrees 2 by omitting one of the variables
and using 5 instead of 13. In this section, an algorithm that
helps in factorizing the expression to find the maximum number of
common factros will be presented. Experimental results for that
algorithm will be shown in the next section. To factorize an
expression maximizing the number of factors, we follow the
following algorithm:

\begin{itemize}

\item[1-] Group the Boolean expression to a collection of minterms
of the same degree and apply the following steps separately on
each collection.

\item[2-] For a homogeneous Boolean function on $n$ variables and
$m$ minterms, draw a table of $n$ columns and $m$ rows such that
each column represents a variable and each row represents a
minterm.

\item[3-] For each minterm in the expression (row), if $x_i$
exists then place 1 in the corresponding cell.

\item[4-] Count the number of 1's in each column and choose the
variable with maximum occurrence as a common factor, if more than
one variable have the same count, then choose the first one. We
will call such variable as {\it a factor variable} and the terms
in the corresponding bracket as {\it a factor group}.

\item[5-] Repeat Step 3 recursively for the remaining rows of the
table collecting a set of factor variables and the corresponding
factor groups, where some minterms will be left over (remainder
minterms).

\item[6-] Find the common terms within all the factor groups and
take them as a common factor generating groups of factor
variables. This step arise the form shown in Observation
\ref{obs1}.

\item[7-] Group the common terms with the same group of factor
variables and repeat Step 3 recursively within each group.

\item[8-] Find the subset relationships between the groups of
factor variables. This arises the form shown in Observation
\ref{obs21}.

\item[9-] For each set of related groups, construct the reversible
circuit as shown in Observation \ref{obs21}. Taking into account
to reset the used variables before constructing the next related
group.

\item[10-] For the remainder minterms, construct the corresponding
$CNOT$ gates using the direct method.
\end{itemize}

For example, consider the Boolean function shown in Eqn.
\ref{HBFR}. The factorization table from steps 1-3 on the
expression is shown in Table.\ref{BentTable}. In this case, all
the variables occurs 8 times in the expression, so we choose $x_0$
as the first factor variable. Applying Step 4 recursively on the
remaining rows, we can find that $x_2$ and $x_5$ are the remaining
factor variables. Next thing to do is to find the common terms in
the factor groups as follows,

\begin{equation}
\begin{array}{l}
 { f = x}_{ 0} ({ x}_{ 1} { x}_{ 2}  \oplus \underline{\underline{\underline {{ x}_{ 1} { x}_{ 3} }}}  \oplus \underline{\underline {{ x}_{ 1} { x}_{ 4} }}  \oplus { x}_{ 1} { x}_{ 5}  \oplus { x}_{ 2} { x}_{ 3}  \oplus { x}_{ 2} { x}_{ 5}  \oplus \underline {{ x}_{ 3} { x}_{ 4} }  \oplus {  }\mathop {{ x}_{ 4} { x}_{ 5} }\limits_{ -  -  -  } ) \\
 \,\,\,\,\,\,\,\, \oplus { x}_{ 2} (\underline{\underline{\underline {{ x}_{ 1} { x}_{ 3} }}} {  } \oplus {  }\underline{\underline {{ x}_{ 1} { x}_{ 4} }}  \oplus \underline {{ x}_{ 3} { x}_{ 4} } {  } \oplus {  x}_{ 3} { x}_{ 5} {  } \oplus {  }\mathop {{ x}_{ 4} { x}_{ 5} }\limits_{ -  -  -  }  ) \\
 \,\,\,\,\,\,\,\, \oplus { x}_{ 5} (\underline{\underline{\underline {{ x}_{ 1} { x}_{ 3} }}} {  } \oplus \underline{\underline {{ x}_{ 1} { x}_{ 4} }}  \oplus \underline {{ x}_{ 3} { x}_{ 4} } ). \\
 \end{array}
\end{equation}

\begin{table}[H]
\begin{center}
\begin{tabular}{|c|c|c|c|c|c|}
\hline
$x_{0}$  & $x_{1}$& $x_{2}$ & $x_{3}$ & $x_{4}$ & $x_{5}$  \\\hline
1&  1&  1&  &   &   \\ \hline
1&  1&  &   1&  &   \\ \hline
1&  1&  &   &   1&  \\ \hline
1&  1&  &   &   &   1\\ \hline
1&  &   1&  1&  &   \\ \hline
1&  &   1&  &   &   1\\ \hline
1&  &   &   1&  1&  \\ \hline
1&  &   &   &   1&  1\\ \hline
&   1&  1&  1&  &   \\ \hline
&   1&  1&  &   1&  \\ \hline
&   1&  &   1&  &   1\\ \hline
&   1&  &   &   1&  1\\ \hline
&   &   1&  1&  1&  \\ \hline
&   &   1&  1&  &   1\\ \hline
&   &   1&  &   1&  1\\ \hline
&   &   &   1&  1&  1\\ \hline

8&  8&  8&  8&  8&  8\\ \hline
\end{tabular}
\caption{Factorization table representing a homogeneous Bent
function of degree 3 on 6 variables.} \label{BentTable}
\end{center}
\end{table}

Then group all the factor terms together follows,

\begin{equation}
\begin{array}{l}
 f = (x_1 (x_3  \oplus x_4 ) \oplus x_3 x_4 )(x_0  \oplus x_2  \oplus x_5 ) \\
  \,\,\,\,\,\,\,\,\oplus x_4 x_5 (x_0  \oplus x_2 ) \\
  \,\,\,\,\,\,\,\,\oplus x_0 x_2 (x_1  \oplus x_3  \oplus x_5 ) \\
  \,\,\,\,\,\,\,\,\oplus x_0 x_1 x_5  \oplus x_2 x_3 x_5.  \\
 \end{array}
\end{equation}

It is clear from the last expression that $(x_0 x_1 x_5  \oplus x_2 x_3 x_5)$ are the remainder minterms.
The group of factor variables $(x_0  \oplus x_2)$ are subset from $(x_0  \oplus x_2  \oplus x_5 )$. Then construct
the reversible circuit as shown in Fig.\ref{BFv6d3}.

\begin{center}
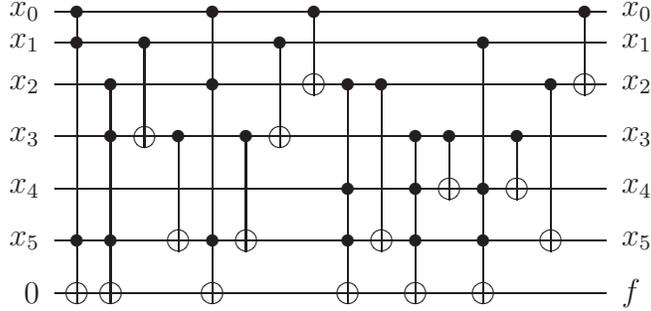
\begin{figure} [htbp]
\begin{center}
\[
\Qcircuit @C=.3em @R=1em {
\lstick{x_0}    & \ctrl{6}  &\qw    &\qw    &\qw    & \ctrl{6}  &\qw    &\qw    & \ctrl{2}  &\qw    &\qw    &\qw    &\qw    &\qw    &\qw    &\qw    & \ctrl{2}  &\rstick{x_0}\qw \\
\lstick{x_1}    & \ctrl{5}  &\qw    & \ctrl{2}  &\qw    &\qw    &\qw    & \ctrl{2}  &\qw    &\qw    &\qw    &\qw    &\qw    & \ctrl{5}  &\qw    &\qw    &\qw    &\rstick{x_1}\qw \\
\lstick{x_2}    &\qw    & \ctrl{4}  &\qw    &\qw    & \ctrl{4}  &\qw    &\qw    & \targ & \ctrl{4}  & \ctrl{3}  &\qw    &\qw    &\qw    &\qw    & \ctrl{3}  & \targ &\rstick{x_2}\qw \\
\lstick{x_3}    &\qw    & \ctrl{3}  & \targ & \ctrl{2}  &\qw    & \ctrl{2}  & \targ &\qw    &\qw    &\qw    & \ctrl{3}  & \ctrl{1}  &\qw    & \ctrl{1}  &\qw    &\qw    &\rstick{x_3}\qw \\
\lstick{x_4}    &\qw    &\qw    &\qw    &\qw    &\qw    &\qw    &\qw    &\qw    & \ctrl{2}  &\qw    & \ctrl{2}  & \targ & \ctrl{2}  & \targ &\qw    &\qw    &\rstick{x_4}\qw \\
\lstick{x_5}    & \ctrl{1}  & \ctrl{1}  &\qw    & \targ & \ctrl{1}  & \targ &\qw    &\qw    & \ctrl{1}  & \targ & \ctrl{1}  &\qw    & \ctrl{1}  &\qw    & \targ &\qw    &\rstick{x_5}\qw \\
\lstick{0}  & \targ & \targ &\qw    &\qw    & \targ &\qw    &\qw    &\qw    & \targ &\qw    & \targ &\qw    & \targ &\qw    &\qw    &\qw    &\rstick{f}\qw
}
\]
\end{center}
\caption{Reversible circuit realization of a homogeneous Bent
function of degree 3 on 6 variables shown in Eqn. \ref{HBFR} using
the factorization algorithm .} \label{BFv6d3}
\end{figure}
\end{center}

\subsection{Experimental Results}

Even though the art of synthesis and optimization of reversible
circuits has been there for sometimes
\cite{practmethod,transrules,revlogic,wav}, there are none of the
work has been done focused on the construction of reversible
circuits for homogeneous Boolean functions where some of the
mentioned work couldn't optimize such circuits because of the
nonlinearly of that functions. The proposed algorithm has been
examined on the known benchmarks $2of5$ and $4mod5$ \cite{Benurl2}
that can be represented as a set of homogeneous Boolean functions
of degree at most 3, for example, the Boolean expression of the
$4mod5$ is as follows,

\begin{equation}
\begin{array}{l}
 f = 1 \oplus x_0  \oplus x_1  \oplus x_2  \oplus x_3  \\
  \,\,\,\,\,\,\oplus x_0 x_1  \oplus x_1 x_2  \oplus x_0 x_3  \oplus x_2 x_3,  \\
 \end{array}
\end{equation}

\noindent

where it contains 5 minterms of degree 1 and 4 minterms of degree
2, applying the factorization algorithm, the synthesized circuit
is shown in Fig. \ref{4mod5F}, where the last three $CNOT$ gates
can be removed if the state of the inputs is not important.

The algorithm also has been tested on representatives of
homogenous bent functions of degree 3 with 6 and 8 variables. It
was shown in \cite{Benur3} a set of 30 homogeneous bent functions
of degree 3 with 6 variables each with 16 minterms ($BFV6{\_}d3$),
and a set of 20 homogeneous bent functions of degree 3 with 8
variables, 4 functions with 24 minterms ($BFV8{\_}d3{\_}24$), 6
functions with 28 minterms ($BFV8{\_}d3{\_}28$), 4 functions with
32 minterms ($BFV8{\_}d3{\_}32$) and 6 functions with 35 minterms
($BFV8{\_}d3{\_}35$). The experimental results is shown in Table
\ref{tab1}. The results are shown by omitting the $CNOT$ gates
that restore the state of the inputs to follow the known
benchmarks for $2of5$ and $4mod5$\cite{Benurl2}. Even if the
missing $CNOT$ gates are added to the results, there will be no
increase in the number of $CNOT$ gates compared with the direct
synthesis method. The results show an improvement in the quantum
cost compared with the best known results in the literature. It
can be seen from the results of bent functions of 8 variables that
the degree of improvement increases as the number of minterms
increases.

\begin{center}
\begin{figure} [htbp]
\begin{center}
\[
\Qcircuit @C=.3em @R=1em {
\lstick{x_0}    &\qw    & \ctrl{2}  &\qw    &\qw    &\qw    &\qw    &\qw    &\qw    & \ctrl{2}  &\rstick{x_0}\qw \\
\lstick{x_1}    &\qw    &\qw    & \ctrl{2}  &\qw    &\qw    &\qw    &\qw    & \ctrl{2}  &\qw    &\rstick{x_1}\qw \\
\lstick{x_2}    &\qw    & \targ &\qw    & \ctrl{2}  & \ctrl{1}  &\qw    & \ctrl{1}  &\qw    & \targ &\rstick{x_2}\qw \\
\lstick{x_3}    &\qw    &\qw    & \targ & \ctrl{1}  & \targ & \ctrl{1}  & \targ & \targ &\qw    &\rstick{x_3}\qw \\
\lstick{0}  & \targ &\qw    &\qw    & \targ &\qw    & \targ &\qw
&\qw    &\qw    &\rstick{f}\qw}
\]
\end{center}
\caption{Reversible circuit realization of the $4mod5$ function
using the factorization algorithm.} \label{4mod5F}
\end{figure}
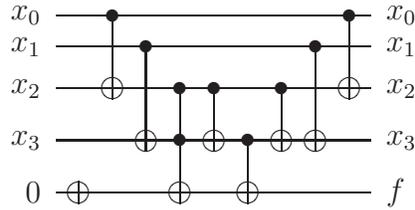
\end{center}

\begin{table}[htbp]

\begin{tabular}
 {|c|c|c|c|c|c|c|c|}
\hline \raisebox{-1.50ex}[0cm][0cm]{Benchmark}&
\raisebox{-1.50ex}[0cm][0cm]{Inputs}&
\multicolumn{2}{|c|}{Bechmark\cite{Benurl2}} &
\multicolumn{2}{|c|}{Direct Synthesis} &
\multicolumn{2}{|c|}{Factorization}  \\
\cline{3-8}
 &
 &
Gates&
Cost&
Gates&
Cost&
Gates&
Cost \\
\hline $4mod5$& 4& 5& 13& 9& 25& 6&
8 \\
\hline $2of5$& 5& 15& 107& 20& 180& 17&
75 \\
\hline BFV6{\_}d3& 6& - & -& 16& 208& 13&
85 \\
\hline $BFV8{\_}d3{\_}24$& 8& -& -& 24& 312& 24&
196 \\
\hline $BFV8{\_}d3{\_}28$& 8& -& -& 28& 364& 28&
184 \\
\hline $BFV8{\_}d3{\_}32$& 8& -& -& 32& 416& 32&
188 \\
\hline $BFV8{\_}d3{\_}35$& 8& -& -& 35& 455& 32&
176 \\
\hline
\end{tabular}
\caption{Reversible circuits bechmarks, where $-$ indicates that results are not available.}
\label{tab1}
\end{table}

\section{Conclusion}
\label{sec4}

This paper shown a simple algorithm based on factorization of the
algebraic expression of a homogeneous Boolean function of degree
at most 3. The algorithm finds the common parts of the circuit and
decrease the number of their computation, this will decrease the
quantum cost of the generated circuit.

The algorithm can also perform on a Boolean function that can be
divided into a group of homogeneous Boolean functions. The ability
of the algorithm to minimize the quantum cost increases as the
number of minterms of the same degree increases.

\bibliography{bentf3}
\bibliographystyle{plain}

\end{document}